# Time-dependent Mandel Q parameter analysis for a hexagonal boron nitride single photon source

Callum Jones,[1,*] Jolly Xavier,[1,2] Samir Vartabi Kashanian,[1] Minh Nguyen,[3] Igor Aharonovich,[3,4] and Frank Vollmer[1]

[1]*Living Systems Institute, University of Exeter, Stocker Road, Exeter, EX4 4QD, United Kingdom*
[2]*Currently with SeNSE, Indian Institute of Technology Delhi, Hauz Khas, New Delhi, Delhi, India*
[3]*School of Mathematical and Physical Sciences, Faculty of Science, University of Technology Sydney, Ultimo, New South Wales, 2007, Australia*
[4]*ARC Centre of Excellence for Transformative Meta-Optical Systems (TMOS), University of Technology Sydney, Ultimo, New South Wales, 2007, Australia*
*\*cj403@exeter.ac.uk*

**Abstract:** The time-dependent Mandel Q parameter, $Q(T)$, provides a measure of photon number variance for a light source as a function of integration time. Here, we use $Q(T)$ to characterise single photon emission from a quantum emitter in hexagonal boron nitride (hBN). Under pulsed excitation a negative Q parameter was measured, indicating photon antibunching at an integration time of 100 ns. For larger integration times Q is positive and the photon statistics become super-Poissonian, and we show by comparison with a Monte Carlo simulation for a three-level emitter that this is consistent with the effect of a metastable shelving state. Looking towards technological applications for hBN single photon sources, we propose that $Q(T)$ provides valuable information on the intensity stability of single photon emission. This is useful in addition to the commonly used $g^{(2)}(\tau)$ function for the complete characterisation of a hBN emitter.

## 1. Introduction

Hexagonal boron nitride (hBN) is a two-dimensional wide bandgap semiconductor material which hosts atomic vacancy defects. These defects are now widely studied because they can produce bright single photon emission even at room temperature [1, 2], with the possibility of being integrated into two-dimensional heterostructure devices [3, 4]. Single photon sources based on hBN are beginning to be applied in quantum information technologies, for example in quantum key distribution (QKD) [5, 6], fundamental tests of quantum theory [7], as well as integrating quantum emitters with optical cavities [8, 9, 10, 11].

A key component of quantum optical technologies is the single photon on demand source: a source which can deliver deterministic single photon pulses at high count rates [12, 13]. These sources may find applications in QKD [14], optical quantum computing, radiometry [15, 16], or even in probing the single photon response of biological systems [17]. For a quantum emitter such as those in hBN to be used as a single photon on demand source for these applications, it must have a photon number distribution as close as possible to the single photon number state. In other words, both multiphoton and zero photon pulses must be suppressed so that the photon number variance is reduced. The range of timescales over which this small photon number variance is preserved is a measure of the intensity stability of the source. Quantifying this stability is our motivation to measure the time-dependent Mandel Q parameter for a quantum emitter in hBN.

The time-dependent Mandel Q parameter is defined as the variance over mean minus one for the photon number $N$ measured from a source during integration time $T$ [18, 13, 19]:

$$Q(T) = \frac{\langle \Delta N^2 \rangle_T}{\langle N \rangle_T} - 1 \,. \tag{1}$$

$\langle \ldots \rangle_T$ denotes averaging over the integration time $T$. The photon number for a coherent state follows a Poisson distribution, therefore $\langle \Delta N^2 \rangle_T = \langle N \rangle_T$ and $Q(T) = 0$. The expected result for an ideal single photon source is a negative Q parameter, which indicates the photon number distribution is narrower than a Poisson distribution and so is an indication of intensity squeezing.

In previous works $Q(T)$ has been used to characterise single photon emission from a single molecule triggered single photon source [13]. Other more varied applications outside quantum information technologies include the use of $Q(T)$ for single molecule measurements, for example the detection of singlet oxygen or the fast recognition of single dye molecules [20, 21]. We note that the abstracts in Refs. [22, 23] mention observations of negative $Q(T)$ on timescales of ~10 ns for hBN emitters, therefore this topic demands a full study.

In this paper we will present the characterisation of a stable quantum emitter in hBN, focusing on the time-dependent Mandel Q parameter. With the help of data simulated by Monte Carlo methods, we show that the effect of the detector deadtime prevented us from observing antibunching in the Q parameter under CW excitation. However, for pulsed excitation with appropriate parameters, a negative Q parameter was found when integrating over one pulse period (100 ns). Finally, we discuss the use of $Q(T)$ to complement other measures in the complete characterisation of photon statistics for hBN quantum emitters.

## 2. Methods and Materials

### 2.1 Experimental setup

Solvent-exfoliated multilayer hBN nanoflakes (Graphene Supermarket) were dropcast onto a 5×5 mm Si/SiO$_2$ substrate, followed by hotplate annealing at 500°C to remove carbonaceous material. The quantum emitters were then activated by thermal annealing at 850°C under an Ar atmosphere.

Samples prepared in this way were transferred to a custom-built confocal microscope in order to study photoluminescence (PL) from individual quantum emitters. The setup uses either continuous wave (CW) (Prometheus, Coherent Inc.) or picosecond pulsed (PicoQuant, pulse length 72 ps) lasers at 532 nm. The input beam is focused through a 100x (NA 0.9) objective onto the sample mounted on an XYZ piezo stage (PI Ltd.).

PL from the sample is collected through the same objective and separated from the pump beam by two dichroic mirrors (transmitting > 550 nm). The final stage of the setup is spectral filtering: first a long-pass interference filter transmitting > 568 nm (Semrock) removes the remaining 532 nm beam, then an angle tunable band-pass filter with a 20 nm bandwidth tunable from 560-630 nm selects the spectral region with the highest PL intensity. The output light is collected into a multimode fibre (50 μm core diameter). To prevent misalignment with the fibre coupler when the tunable filter is rotated, a compensating plate was added after the filter which is calibrated to rotate in the opposite direction to the filter and compensate the beam walk-off from the optic axis.

A multimode fibre beamsplitter splits the output light onto two single photon avalanche diodes (SPADs, MPD), in the Hanbury Brown and Twiss (HBT) configuration. This experimental configuration was used for all the following measurements.

### 2.2 Single photon emitter characterisation

Individual quantum emitters were located by scanning the sample piezo stage to produce PL maps of the surface under CW excitation, see Fig. 1b. All the following measurements were carried out on one emitter which was exceptionally stable. No blinking was observed for this emitter under any excitation power we used over hours of measurement. Fluorescence blinking on the order of seconds has been reported to change the transition rates between emitter energy levels [24]; we expect our emitter to be free from this effect. The blinking we refer to here is a transition between a bright and dark state of the emitter (common to the literature on hBN [24, 2]), not the transition to a long-lived shelving state as the term blinking is used in Ref. [13].

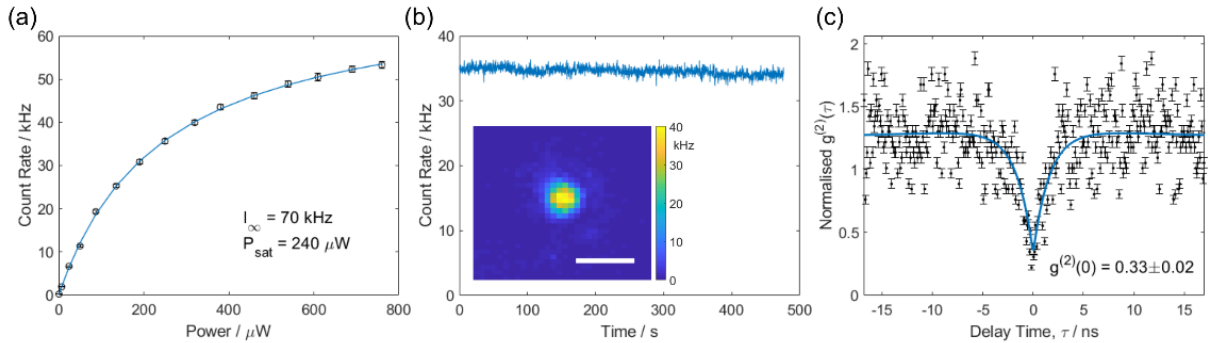

**Fig.1.** (a) Photoluminescence (PL) count rate saturation curve for our hBN quantum emitter. The saturation power $P_{sat}$ is 240 μW. (b) PL count rate over time with 250 μW CW excitation showing stable emission rate. The emitter showed no blinking events under any of the conditions studied; this is also shown by the small error bars in the saturation curve. Inset: PL map (photon count rate in Hz) of the quantum emitter. Scale bar: 1 μm. (c) Second order correlation histogram for 250 μW CW excitation. A two-exponential fit was applied to obtain $g^{(2)}(0) = 0.33 \pm 0.02$, confirming single photon emission.

The count rate saturation with increasing pump power is shown in Fig. 1a. Fitting the following function to the saturation curve gave a saturation power of $P_{sat} = 240$ µW [24]:

$$I = \frac{I_\infty P}{P + P_{sat}} + bP + c. \tag{2}$$

$I$: count rate, $I_\infty$: max. count rate, $b$: power dependent background, and $c$: constant background. The beam diameter at the sample surface was measured to be approximately 1 µm.

Single photon emission was confirmed by measuring the second order correlation function $g^{(2)}(\tau)$ with a value of $g^{(2)}(0) = 0.33 \pm 0.02$ for CW excitation at 250 µW, Fig. 1c. A two exponential fit was applied with antibunching/bunching timescales $\tau_1$, $\tau_2$ to obtain the value of $g^{(2)}(0)$ (as in [1], except that we use both antibunching and bunching parameters $A, B$ to account for non-zero values of $g^{(2)}(0)$):

$$g^{(2)}(\tau) = 1 - (1+A)e^{-\frac{|\tau|}{\tau_1}} + Be^{-\frac{|\tau|}{\tau_2}}. \tag{3}$$

The effect of uncorrelated background counts (assumed to be Poisson distributed) was removed for the CW measurements using the signal to background ratio $SBR$ estimated from the PL intensity map of the emitter. For the $SBR$, from which we define $\sigma = SBR/(1+SBR)$ (in Fig.1c, $\sigma = 0.987$), the correction is [24, 25]:

$$g^{(2)}_{corrected}(\tau) = \frac{g^{(2)}_{raw}(\tau) + \sigma^2 - 1}{\sigma^2}. \tag{4}$$

To produce the $g^{(2)}(\tau)$ histogram, photon arrival times on the two SPADs were recorded using a photon counting module (Time Controller, ID Quantique). The time resolution of our setup is limited by the SPAD timing jitter, nominally 50 ps.

### 2.3 Mandel Q parameter calculation

To measure the intensity stability of our emitter, the time-dependent Mandel Q parameter $Q(T)$ was calculated according to Eq. (1). This was calculated using the same raw photon arrival times, or timestamps, used for $g^{(2)}(\tau)$ measurements. Timestamps were collected in acquisition times of 100 s for all measurements.

To calculate $Q(T)$, timestamps from both detectors were combined into one array, sorted, and split into $K$ time windows of length $T$ using a Matlab script. Unless stated otherwise, $K = 10^8$ time windows were used (or the maximum allowed by the 100 s acquisition time for a given $T$) to calculate the variance and mean of the photon number per window $\langle \Delta N^2 \rangle_T$ and $\langle N \rangle_T$, and hence $Q(T)$. For pulsed excitation the $T$ values were all multiples of the pulse repetition rate. Error bars were calculated by repeating the analysis over many acquisitions.

This analysis was carried out on timestamps from both CW (see Section 3.1) and pulsed (see Section 3.2) excitation. In addition to the procedure above, pulsed timestamps were also filtered using the trigger pulse from the laser, as in [13]. Only timestamps occurring within a window starting from the arrival of the laser pulse were kept. The optimum filter width was found to be 5 ns; this is discussed further in the Supplemental Document.

### 2.4 Modelling hBN energy levels

In order to offer a comparison to experimental data for $Q(T)$ and better understand the photodynamics of our hBN emitter, we simulated timestamp data using Monte Carlo methods. We used a three-level system as a simplified model for the emitter, defined by four transition lifetimes (see Fig. 2b). Although other works have found some emitters require four energy levels [24, 26], we found that models with three and four energy levels fit our $g^{(2)}(\tau)$ data equally well.

Timestamps were simulated based on the method detailed in [25]. In each excitation cycle, the four lifetimes for each transition are drawn from exponential distributions to determine whether a photon is emitted (is the cycle radiative or non-radiative), and at what time. Losses are modelled by removing photon detections based on a binomial distribution, the detections are split 50:50 into two channels to model the beamsplitter in the HBT setup, and finally the detector deadtime is modelled by removing any detections less than the deadtime (80 ns for our SPADs) after the previous detection. The resulting timestamps are in an identical format to our experimental data, so the same data analysis for $g^{(2)}(\tau)$ and $Q(T)$ was applied.

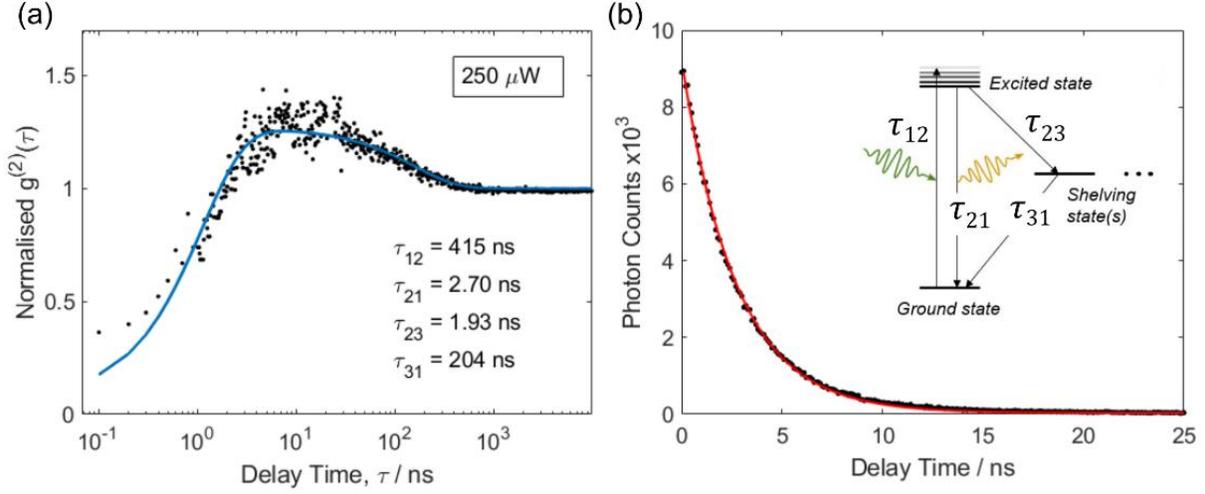

**Fig.2.** (a) Fitting a three-level model to g$^{(2)}$ data for CW excitation at 250 µW. The best fit transition lifetimes $\tau_{ij}$ are shown; lifetimes for the 540 and 760 µW are in Table 1. (b) Radiative lifetime measurement made using pulsed excitation at 10 MHz and 24 µW mean power; data is shown for one 100 s acquisition. Fitting a single exponential to all 144 acquisitions of timestamp data gives a mean value of $\tau_{21} = 2.7 \pm 0.1$ ns. Inset: Schematic of the three-level model for the emitter showing the transition lifetimes $\tau_{ij}$ between levels.

To run this simulation we needed to know appropriate transition lifetimes to set as mean values in the Monte Carlo model. We determined these parameters by fitting the three-level model to experimental $g^{(2)}(\tau)$ data plotted on a log scale over the time interval 100 ps – 10 µs by a procedure described in [24]. For a three-level system, the populations of the energy levels $\rho_i(t)$ are described by a system of coupled differential equations:

$$\frac{d\rho_1(t)}{dt} = -k_{12}\rho_1(t) + k_{21}\rho_2(t) + k_{31}\rho_3(t)$$
$$\frac{d\rho_2(t)}{dt} = k_{12}\rho_1(t) - k_{21}\rho_2(t) - k_{23}\rho_2(t) \qquad (5)$$
$$\frac{d\rho_3(t)}{dt} = k_{23}\rho_2(t) - k_{31}\rho_3(t).$$

The transition rates $k_{ij}$ are used as fitting parameters for this model, which is related to the normalised $g^{(2)}(\tau)$ by the solution to the differential equations for $\rho_2(\tau)$, with the initial conditions $\rho_1(0) = 1$, $\rho_{2,3}(0) = 0$ [27]:

$$g^{(2)}(\tau) = \frac{\rho_2(\tau)}{\lim_{t \to \infty} \rho_2(t)}. \qquad (6)$$

This model will always have $g^{(2)}(0) = 0$, therefore to include the effect of the background counts we used the background correction Eq. (4) to include $\sigma$ as a fitting parameter. The model was fitted to experimental data using the parameters $k_{ij}$, which are the reciprocals of the transition lifetimes $\tau_{ij}$ represented in Fig. 2b.

**Table 1. Fitting parameters for $g^{(2)}(\tau)$ data**

| Power / µW | $\tau_{12}$ / ns | $\tau_{21}$ / ns | $\tau_{23}$ / ns | $\tau_{31}$ / ns | $\sigma$ |
|---|---|---|---|---|---|
| 250 (1.0 $P_{sat}$) | 415 | 2.70 | 1.93 | 204 | 0.962 |
| 540 (2.3 $P_{sat}$) | 189 | 2.83 | 1.85 | 195 | 0.722 |
| 760 (3.2 $P_{sat}$) | 135 | 2.86 | 1.81 | 125 | 0.811 |

Additional constraints were needed to fit the four parameters $\tau_{ij}$: the radiative lifetime, $\tau_{21}$, was measured directly using pulsed excitation at 10 MHz and 24 µW mean power. A single exponential fit to the lifetime curve in Fig. 2b gave $\tau_{21} = 2.7 \pm 0.1$ ns. In addition, the excitation lifetime $\tau_{12}$ was assumed to be linearly proportional to the power (as in [24]) and measurements at 250, 540 and 760 µW were used to constrain $\tau_{12}$. The best fit parameters $\tau_{ij}$ and $\sigma$ at the three powers are given in Table 1, these were used as the initial parameters for the Monte Carlo simulated timestamps.

## 3. Results and Discussion

### 3.1 Continuous-wave Mandel Q parameter

First, we examined the photon number distribution of our emitter for 100 s of timestamp data under CW excitation at 250 µW, see Fig. 3. The single photon emission follows a distribution very close to a Poisson distribution over integration times from 1 µs to 1 ms. The photon number variance across these integration times is therefore almost the same as that of a coherent state. Since $g^{(2)}(\tau)$ clearly shows antibunching over ns timescales (e.g. in Fig. 1c, 2a), we would expect that as the integration time is decreased further the photon number variance will become significantly smaller than that of a Poisson distribution. The difference between the data and Poisson distribution variances is better visualized over several orders of magnitude in integration time using the Mandel Q parameter $Q(T)$.

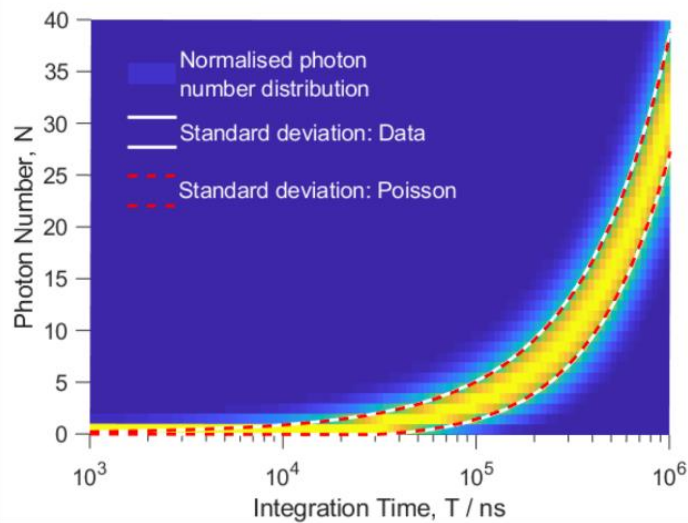

**Fig.3.** Photon number distribution for CW emission under 250 µW excitation. Along the y axis is a histogram of photon number per time bin, as a function of the time bin width or integration time, T. Each histogram height is normalised to one so the widths of the distributions can be compared. Curves indicate one standard deviation for the data (white), and a Poisson distribution with the same mean photon number (red). The data and Poisson deviations overlap almost exactly on this scale.

The Mandel Q parameter was calculated for CW excitation at three powers: 250, 540 and 760 µW, see Fig. 4a. We see that above approximately 100 ns the photon number distribution is wider than a Poisson distribution. Below this integration time there is a small negative value; the minimum is $Q = -(1.4 \pm 0.3) \times 10^{-3}$ for 540 µW, but $Q(T)$ approaches zero at lower $T$. The measurement is limited at low integration times because the mean photon number per time bin tends to zero. At integration times $\ll 1/I$, where $I$ is the emitter count rate, most time bins contain no photons. This is also illustrated by Fig. 3. For 250 µW excitation, $I = (34 \pm 3)$ kHz, therefore the average photon number is less than one below $T = 29 \pm 3$ µs.

$Q(T)$ for simulated timestamp data can reproduce a similar form to the experimental data using the three-level Monte Carlo model described in Section 2.4. In Fig. 4b we compare the result for a two-level and three-level emitter, in which the transition lifetimes for the first two levels are kept the same (red and blue points, respectively). The transition lifetimes found in Table 1. were optimized to match the simulated $Q(T)$ more closely to experiment and to have the same acquisition time as the experimental data. The final simulation parameters were: $\tau_{12} = 205$ ns, $\tau_{21} = 1.60$ ns, $\tau_{23} = 1.40$ ns, $\tau_{31} = 420$ ns, and $\eta_{model} = 0.248$. The two-level model uses only $\tau_{12}$ and $\tau_{21}$, and $\eta_{model}$ is the fraction of photons detected in the model.

Fig. 4b illustrates that the positive $Q(T)$ values are a consequence of adding a metastable shelving state. For a two-level emitter with only a ground and excited state (red points), the antibunched photon statistics and negative $Q(T)$ are maintained as the integration time is increased. However, adding a shelving state (blue points) introduces a non-radiative decay path causing fluctuations in the single photon count rate. These become significant for integration times higher than the shelving state lifetime, which was $\tau_{31} = 420$ ns in this case.

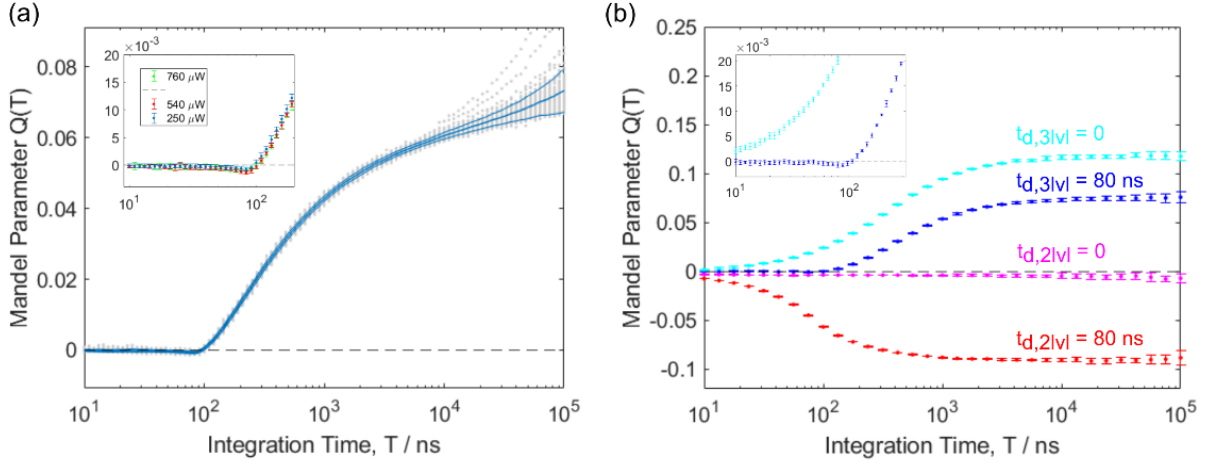

**Fig.4.** (a) Experimental Mandel Q(T) for CW excitation at 250 µW. Blue lines show mean values over 99 separate 100 s acquisition times, plus and minus one standard deviation. Inset: zoom-in showing small negative values for all three powers 250, 540 and 760 µW. (b) Simulated CW Mandel Q(T) from Monte Carlo model. Blue (cyan) points were simulated using a three-level model with (without) modelling the 80 ns detector deadtime $t_d$. Red (magenta) points are from a two-level model with (without) the detector deadtime. In both cases removing the deadtime effect makes all Q(T) values more positive, and for the three-level simulation removing the deadtime removes all negative values (inset). Comparing the two- and three-level simulations we can see that adding a metastable shelving state causes Q(T) to become positive at long integration times.

An important factor that must be considered is the detector deadtime. When the deadtime is artificially removed from the simulated timestamp data, $Q(T)$ becomes positive for all integration times, i.e. the negative value disappears (cyan points in Fig. 4b). The deadtime introduces an artificial antibunching effect since there is a decreased probability of two photon detections within the deadtime period of around 80 ns. Therefore, *the negative points in Fig. 4a are likely due to the detector deadtime only*, especially considering the minimum value of $Q(T)$ occurs near 80 ns. As discussed in Refs. [13, 28], pulsed excitation with a repetition period longer than the deadtime must be used to avoid the deadtime artefact.

### 3.2 Pulsed Mandel Q parameter

For pulsed excitation, a 10 MHz repetition rate was used so that the repetition period of 100 ns was longer than the detector deadtime of 80 ns. A mean power of 24 µW (pulse energy 2.4 pJ) produced a $g^{(2)}(0)$ value of $0.37 \pm 0.02$ (Fig. 5a) with a count rate of 2.8 kHz; increasing the power to 82 µW and 160 µW increased $g^{(2)}(0)$ above 0.5. The timestamps were filtered using a trigger pulse from the laser: only timestamps within 5 ns of the pulse start were used, i.e. approx. 2× the radiative lifetime. Four hours of timestamps at 24 µW measured in 100 s acquisitions were used to produce the $g^{(2)}(\tau)$ and $Q(T)$ plots in Fig. 5.

We found a negative value at $T = 100$ ns of $Q = -(1.4 \pm 0.8) \times 10^{-4}$. As for the CW case, $Q(T)$ is positive for integration times greater than 100 ns. This negative value is very small; the magnitude of $Q(T)$ depends on the total detection efficiency of the measurement. The expected value for integration over one pulse period is [28]:

$$Q_0(T) = \eta \left( \frac{g^{(2)}(0)}{2} - 1 \right) \qquad (7)$$

where $\eta$ is the total photon detection efficiency (i.e. including both the collection efficiency of photons and the detector efficiency). From our experimental values we can estimate $\eta = (2 \pm 1) \times 10^{-4}$. To measure a more negative value and a stronger signature of photon antibunching would require both a higher total detection efficiency and lower $g^{(2)}(0)$. We note that several works on hBN emitters have achieved low $g^{(2)}(0)$ values (as low as 0.07) under pulsed excitation [3, 29, 6, 5]. In our experiments we found a difficult trade-off in the pulsed excitation power between low $g^{(2)}(0)$ improved by low power and achieving a sufficiently high count rate to make our measurements.

Simulated timestamp data was compared with or without the deadtime modelled in Fig. 5c. The data were simulated with the method described in Section 2.4 using the parameters: $\tau_{12} = 100$ ps, $\tau_{21} = 2.70$ ns, $\tau_{23} = 2.40$ ns, $\tau_{31} = 420$ ns, and total detection efficiency $\eta_{model} = 2.54 \times 10^{-3}$.

There is no significant difference in $Q(T)$ when the deadtime is removed for pulsed excitation at 10 MHz, and crucially, the negative value at $T = 100$ ns remains. Therefore, we conclude that the experimental $Q(T)$ data for 24 µW pulsed excitation shows evidence of antibunching at $T = 100$ ns and bunching at longer integration times. As in the case of CW excitation, the bunching is attributed to the presence of a metastable shelving state.

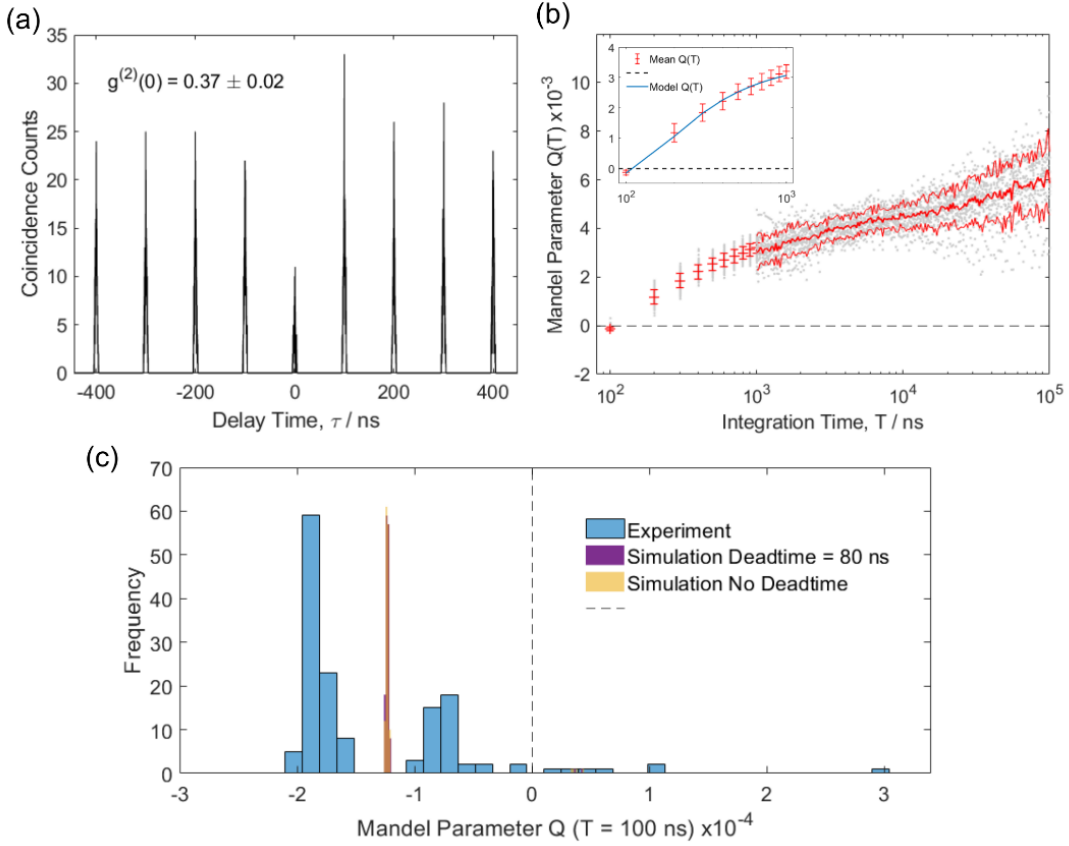

**Fig.5.** (a) Second order correlation function $g^{(2)}(\tau)$ for 10 MHz pulsed excitation at 24 µW mean power. $g^{(2)}(0) = 0.37 \pm 0.02$ was calculated using the relative area of the peak at zero delay time. Note that the timestamp data was filtered keeping only detections within 5 ns of the laser trigger pulse. (b) Mandel Q(T) for pulsed excitation at 24 µW showing positive Q(T) for integration times greater than 100 ns. Red lines show mean values +/- one standard deviation. Inset: fit to data using Eq. (8). (c) Histogram of Q(T) values at T = 100 ns (integrating over one pulse period, blue bars). Values are compared over 144 acquisitions of 100 s each. The mean value is $Q = -(1.4 \pm 0.8) \times 10^{-4}$. The purple and yellow bars are Q(T) histograms from simulated timestamp data with comparable parameters, with/without modelling the detector deadtime. The distribution of Q(T) does not change when the simulated deadtime is removed, confirming that the detector deadtime has no effect on Q(T) when the pulse repetition period is greater than the deadtime.

The lifetime of the shelving state can be estimated by fitting the model from [28] to the pulsed excitation $Q(T)$ data, expressed in terms of the shelving/deshelving lifetimes $\tau_{23}, \tau_{31}$, the integration time $T = k\tau_{rep}$ ($k$ is an integer number of pulse repetition periods $\tau_{rep}$), and the total detection efficiency $\eta$:

$$Q(k\tau_{rep}) = \eta \cdot \left[ \frac{\tau_{31}}{\tau_{23} + \tau_{31}} \left( \frac{2-\beta}{\beta} - \frac{2(1-\beta)}{k} \cdot \frac{1-(1-\beta)^k}{\beta^2} \right) - 1 \right] \qquad (8)$$

where $\beta = \tau_{rep}(1/\tau_{23} + 1/\tau_{31})$.

A simplified form of this model was used to fit data from the single molecule triggered single photon source in [13] by assuming that $\tau_{rep}/\tau_{23} \ll 1$; $\tau_{rep}/\tau_{31} \ll 1$ is satisfied. Our hBN emitter does not meet these criteria:

$\tau_{rep}/\tau_{23} = 0.67$ and $\tau_{rep}/\tau_{31} = 0.14$, therefore we use the full expression Eq. (8) to fit to our $Q(T)$ data (shown in Fig. 5b inset). This yields values of $\tau_{23} = 153$ ns and $\tau_{31} = 665$ ns.

We note that in comparison with the single molecule fluorophore in [13] (having $\tau_{23} = 3.85$ ms and $\tau_{31} = 250$ $\mu$s), the hBN emitter has a much shorter shelving state lifetime. This makes the window of observable negative $Q(T)$ values, between a typical detector deadtime and the transition to $Q(T) > 0$, narrower and more difficult to observe.

The crossover time at which $Q(T) = 0$ is a useful metric as it gives a measure of the stability of a single photon source over long timescales. Our single photon emitter had a crossover between 100 and 200 ns. A perfect single photon source, such as an ideal two-level system, would have $Q(T) < 0$ for all integration times.

For certain applications of hBN single photon sources the photon number distribution will be critical, for example single photon on demand sources or in probing the single photon response of biological systems [17]. In these cases, we need to not only confirm single photon emission using $g^{(2)}(\tau)$, but also quantify the single photon variance over timescales which will be relevant to the chosen application. The Mandel Q parameter provides a measure of the photon variance, provided that artefacts due to the detector deadtime are avoided properly. While the absolute value of $Q(T)$ is influenced by detection efficiency, the crossover time at which the photon statistics transform from sub- to super-Poissonian could be taken as a measure of long-term stability of photon antibunching from a single photon source.

Further improvements to hBN emitters for technological applications could look to improving the stability of emission over longer timescales. Currently, photon bunching seems to be inevitable over ~100 ns timescales due to the presence of a metastable shelving state. However, hBN quantum emitters with different chemical structures have significantly different energy level structures and shelving state lifetimes [30]. An emitter with a longer shelving state lifetime and higher preference for radiative decay should offer better single photon antibunching stability.

## 4. Conclusion

We presented measurements of the time-dependent Mandel Q parameter $Q(T)$ for a quantum emitter in hBN, which showed a very stable count rate and no fluorescence blinking. A comparison between experimental and simulated $Q(T)$ for CW excitation demonstrated that artefacts due to the deadtime of the single photon detectors prevent us from measuring photon antibunching.

By using 10 MHz pulsed excitation instead, with the pulse repetition rate longer than the detector deadtime, a negative value of $Q(T) = -(1.4 \pm 0.8) \times 10^{-4}$ was obtained at $T = 100$ ns. At longer integration times $Q(T)$ becomes positive due to photon bunching caused by the metastable shelving state, an effect we could also demonstrate using the simulated data.

The transition from sub- to super-Poissonian photon statistics occurs between 100 and 200 ns, which is significantly shorter than the crossover time of a few μs for previous measurements on a single fluorescent molecule [13]. There is therefore a narrow window between our detector deadtime of 80 ns and the crossover time in which $Q(T) < 0$ can be observed.

Monte Carlo simulations based on transition rates obtained by fitting to the long-timescale $g^{(2)}(\tau)$ allowed us to simulate timestamps that were comparable to the specific hBN emitter being studied. This method proved useful for investigating the effects of the detector deadtime and metastable shelving state on $Q(T)$ by allowing us to artificially remove these effects.

The Mandel Q parameter provides a measure of the intensity stability of a single photon source, in particular the crossover time indicates the integration time up to which photon antibunching is maintained. This is important for characterizing hBN single photon emitters towards applications in quantum information technologies such as single photon on demand sources.

**Funding.** Engineering and Physical Sciences Research Council (EP/R031428/1); Royal Society (WRMA); Australian Research Council (CE200100010).

**Acknowledgments.** We thank the University of Exeter Physics Workshop team for manufacturing custom parts for the confocal microscope.

**Disclosures.** The authors declare no conflicts of interest.

**Data availability.** Data underlying the results presented in this paper are not publicly available at this time but may be obtained from the authors upon reasonable request.

**Supplemental document.** See Supplement 1 for supporting content.

# Time-dependent Mandel Q parameter analysis for a hexagonal boron nitride single photon source: Supplemental Document

Callum Jones,[1,*] Jolly Xavier,[1,2] Samir Vartabi Kashanian,[1] Minh Nguyen,[3] Igor Aharonovich,[3,4] and Frank Vollmer[1]

[1]*Living Systems Institute, University of Exeter, Stocker Road, Exeter, EX4 4QD, United Kingdom*
[2]*Currently with SeNSE, Indian Institute of Technology Delhi, Hauz Khas, New Delhi, Delhi, India*
[3]*School of Mathematical and Physical Sciences, Faculty of Science, University of Technology Sydney, Ultimo, New South Wales, 2007, Australia*
[4]*ARC Centre of Excellence for Transformative Meta-Optical Systems (TMOS), University of Technology Sydney, Ultimo, New South Wales, 2007, Australia*
*\*cj403@exeter.ac.uk*

**Supplemental Document.**

### S1. Pulsed timestamp filter

A filter was applied to the pulsed timestamp data before analysing the $g^{(2)}(\tau)$ and $Q(T)$ functions. Using the trigger pulse output of the pulsed laser, the time delay between each photon detection and the previous trigger pulse was available for all our data. A histogram of the delay time after the trigger pulse is shown in Fig. S1. The filter was applied by keeping only the detections which arrive within a given time window after the trigger pulse; all filters begin from the peak of the pulse at 7 ns delay. The following describes the process of choosing the optimum filter settings to exclude background noise counts and measure $Q(T)$ due to single photon emission counts only.

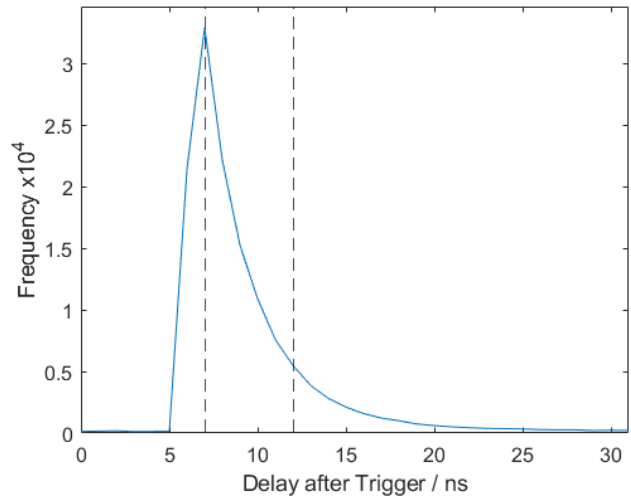

**Fig.S1.** Lifetime curve showing 5 ns filter width applied to the data.

The filter width was varied from 1 ns to applying no filter, i.e. using the raw data. The distribution of Q parameter values for $T = 100$ ns integration time over all 144 data acquisitions is plotted as a function of filter width in Fig. S2. The raw data is plotted at 100 ns filter width as this corresponds to the pulse period, i.e. maximum possible filter width.

As the filter width goes to zero, $Q(T)$ also goes to zero since the number of counts being used for the calculation decreases.

Between 5 ns and 18 ns filter width there is a stable negative $Q(T)$. Above 18 ns however $Q(T)$ increases sharply and becomes positive. There is another increase above 80 ns filter width.

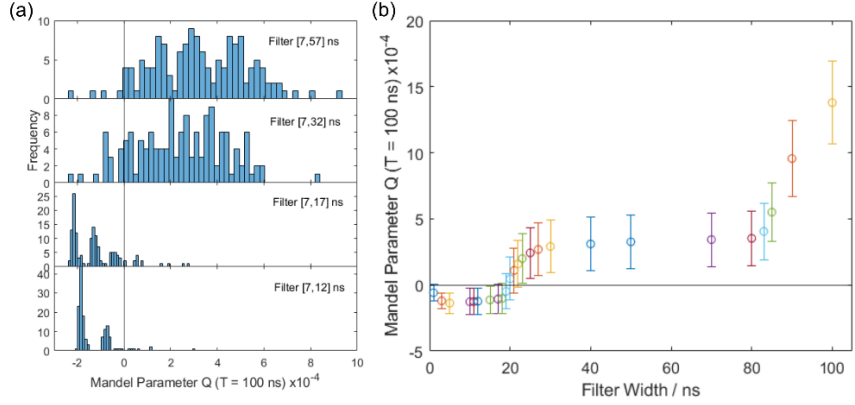

**Fig.S2.** (a) Mandel Q parameter at 100 ns integration time (integrating over one pulse period). Histograms show Q parameter values over the 144 data acquisitions each 100 s long, for different filter widths. (b) Mean Q parameter as a function of filter width.

The sudden increase in $Q(T)$ above 18 ns filter width can be attributed to an artefact of our measurement seen in the raw $g^{(2)}(\tau)$ data, see Fig. S3. There are additional peaks at $\pm 18$ ns delay time in all $g^{(2)}(\tau)$ measurements. This only occurs when using multimode fibre to collect the output light from our setup, the delay time is always the same regardless of the light source being observed, and changing the length of multimode fibre at the output changes the delay time at which the peaks appear. Therefore we conclude these peaks are due to reflections from the end facets of our multimode fibre and as such we treat them as noise.

Using single mode fibres with angled (APC) connectors would solve this issue, however we use multimode fibre, which is widely used in other hBN experiments, in order to collect a high enough count rate under pulsed excitation to perform our measurements.

Note that for CW $g^{(2)}(\tau)$ measurements these peaks were still present, but since the peaks are narrow, they could be excluded from fitting procedures without significantly reducing the number of points used to fit models to the data.

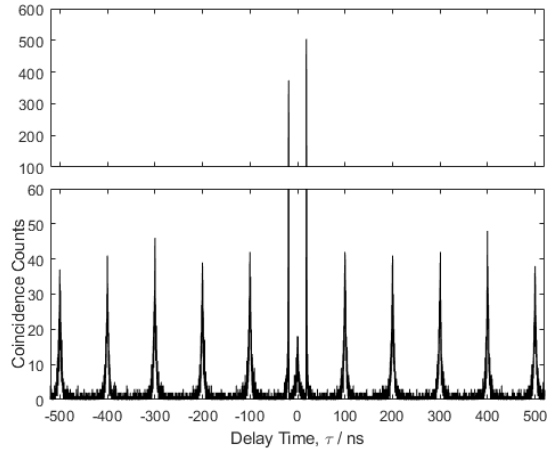

**Fig.S3.** Raw $g^{(2)}(\tau)$ function for pulsed excitation with 24 µW mean power showing noise peaks at $\pm 18$ ns due to reflections in the multimode fibre.

The final choice of filter was 5 ns wide, i.e. over [7,12] ns delay time. This filter setting excludes the noise peaks at $\pm 18$ ns in the $g^{(2)}(\tau)$ function while keeping enough photon counts to measure $Q(T)$ due to single photon counts from our hBN emitter. This filter width is approximately double the radiative lifetime of the emitter: $\tau_{21} = 2.7 \pm 0.1$ ns.

We can also test the effect of applying a filter to the simulated pulsed timestamp data. In Fig. S4a we see that decreasing the filter width below ~5 ns moves $Q(T)$ closer to zero. Unlike the experimental data in Fig. S2b, $Q(T)$ does not change significantly for filter widths above ~5 ns because: a. the simulated data has no background noise, and b. the noise peaks at $\pm 18$ ns delay are not present in the simulated data.

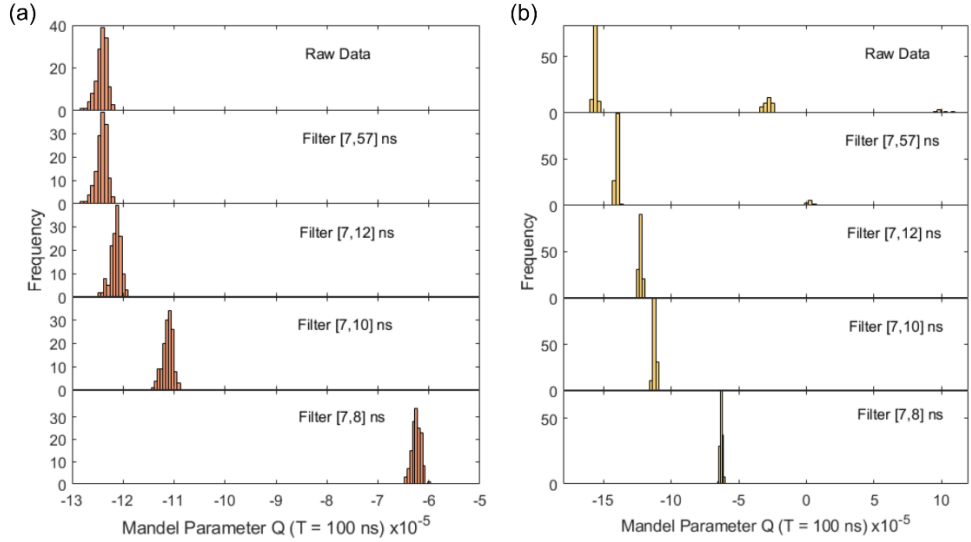

**Fig.S4.** Mandel Q parameter for simulated pulsed timestamp data at 100 ns integration time. (a) Simulated Q parameter as a function of filter width. (b) Simulated Q parameter with uniformly distributed noise counts added to the simulated timestamp data.

One clear difference between the simulated and experimental $Q(T)$ histograms is that the simulation values are all clustered around a single value; the multiple peaks are only seen in the experimental data. We added noise to the simulation by adding uniformly distributed background counts to the simulated timestamps, with the same background count rate of 160±40 Hz per detection channel measured from experimental lifetime curves. With added noise, the simulated $Q(T)$ histograms in Fig. S4b do show multiple peaks and begin to resemble the experimental data more closely.

## S2. Detector deadtime measurement

The deadtime of our single photon avalanche diodes (SPADs) is nominally 77 ns. The deadtime must be known to choose an appropriate pulse repetition rate and to model the deadtime in Monte Carlo simulations. We measured the deadtime by detecting counts from room lights at around 4.4 MHz count rate, approaching the detector saturation count rate. The histogram of time delays between successive counts is shown in Fig. S5.

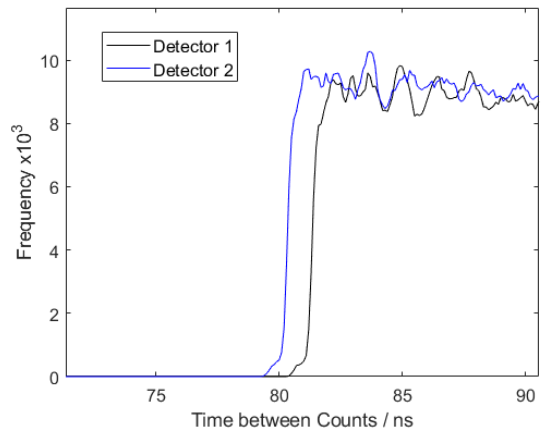

**Fig.S5.** Detector deadtime measurement using a near saturation count rate from room lights. Histogram of delay times between successive photon detections, showing a sudden drop to zero below the detector deadtime. Taking the deadtime as the half rise time of the curves, detector 1 deadtime $t_d = (81.35 \pm 0.10)$ ns; detector 2 deadtime $t_d = (80.35 \pm 0.10)$ ns. At high time delays the histogram slowly decreases to zero because we only consider nearest-neighbour delays.

The histograms show a sudden drop to zero for delays less than the deadtime. At large time delays the histograms slowly decrease to zero because we only accounted for the nearest-neighbour delays. Taking the deadtime to be the half-rise time of the curves gives values of $t_d = (81.35 \pm 0.10)$ ns for detector 1 and $t_d = (80.35 \pm 0.10)$ ns for detector 2. For the purposes of simulating timestamps we took the deadtime to be 80 ns for both detectors.

## S3. Pulsed g$^{(2)}$ as a function of power

Fig. S6 shows the $g^{(2)}(\tau)$ function under pulsed excitation as a function of the mean incident power, at 24 µW, 82 µW and 160 µW. The data for each plot had a 5 ns wide filter applied (over [7,12] ns delay after the trigger pulse). The value $g^{(2)}(0)$ was calculated as the ratio between the $\tau = 0$ peak area and the mean area of the 18 next nearest peaks, and the error was estimated as the standard deviation of the peak areas. Values for $g^{(2)}(0)$ were: $0.37 \pm 0.02$ at 24 µW, $0.69 \pm 0.04$ at 82 µW, and $0.83 \pm 0.06$ at 160 µW.

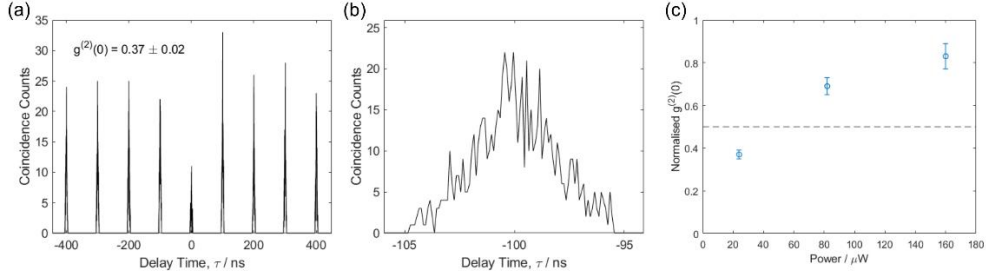

**Fig.S6.** (a) $g^{(2)}(\tau)$ under pulsed excitation at 24 µW. The 5 ns width filter (over [7,12] ns delay after the trigger pulse) was applied to all timestamp data before calculating $g^{(2)}(\tau)$. (b) Zoom-in showing shape of $g^{(2)}(\tau)$ peaks after the filter was applied. (c) Power dependence of $g^{(2)}(0)$.

We found that under pulsed excitation the $g^{(2)}(0)$ value was very sensitive to power and the low power needed to achieve $g^{(2)}(0) < 0.5$ meant that the count rate had to be reduced significantly: the count rate was 2.8 kHz at 24 µW mean power.

The background count rate was measured from a single exponential fit to the 24 µW lifetime curve as 160±40 Hz per detection channel (Fig. 2b in main text). This corresponds to only 0.04 coincidences per time bin in the 24 µW $g^{(2)}(\tau)$ histogram, therefore the background on the pulsed $g^{(2)}$ measurement was ignored. Note that the filtering process does remove background counts occurring outside the filter width.

## S4. Emitter spectral filtering

The output count rate as a function of tunable filter angle was converted into the spectrum in Fig. S7. The tunable filter bandwidth is around 20 nm. All measurements were done with the filter set to the maximum count rate at 595 nm (at 38° to optic axis).

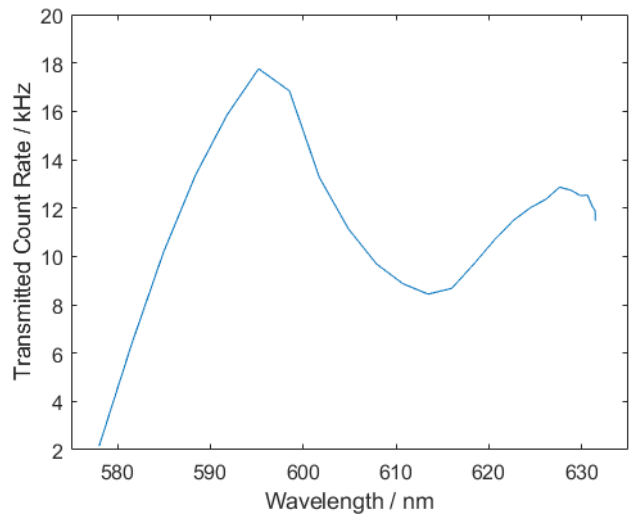

**Fig.S7.** Emission spectrum from our hBN emitter measured by rotating the angle tunable bandpass filter at the output. The bandpass filter has a bandwidth of around 20 nm. Peak emission count rate occurs at 595 nm, corresponding to a filter angle of 38°.

The significance of using spectral filtering is to improve the $g^{(2)}(0)$ value for our emitter: without the filter (filter completely removed from optical path) $g^{(2)}(0) = 0.56 \pm 0.10$, with the filter set to peak count rate $g^{(2)}(0) = 0.33 \pm 0.02$ (see Fig. 1c in main text).

## S5. Analytical solution for CW Q(T)

There is an analytical relation between $g^{(2)}(\tau)$ and $Q(T)$ in the continuous wave (CW) case, for integration time $T$ and average photon count rate $\langle I \rangle$ this is given by [1, 2]:

$$Q(T) = \frac{2\langle I \rangle}{T} \int_0^T d\tau \int_0^\tau d\tau' \left( g^{(2)}(\tau') - 1 \right). \quad (S1)$$

We can calculate this for a $g^{(2)}(\tau)$ function which is well described by a two-exponential fit with lifetimes $t_1, t_2$:

$$g^{(2)}(\tau') = 1 - (1+a)e^{-\frac{|\tau'|}{t_1}} + ae^{-\frac{|\tau'|}{t_2}}. \quad (S2)$$

For simplicity the bunching amplitude is described by one parameter $a$, so that $g^{(2)}(0) = 0$.

Substituting into Eq. (S1):

$$Q(T) = \frac{2\langle I \rangle}{T} \int_0^T d\tau \int_0^\tau d\tau' \left( -(1+a)e^{-\frac{|\tau'|}{t_1}} + ae^{-\frac{|\tau'|}{t_2}} \right)$$

$$Q(T) = \frac{2\langle I \rangle}{T} \int_0^T d\tau \left( -t_1(1+a)\left(1 - e^{-\frac{\tau}{t_1}}\right) + t_2 a \left(1 - e^{-\frac{\tau}{t_2}}\right) \right)$$

$$Q(T) = \frac{2\langle I \rangle}{T} \Big( t_1^2(1+a) - t_2^2 a - (t_1(1+a) - t_2 a)T \\ - t_1^2(1+a)e^{-\frac{T}{t_1}} + t_2^2 a e^{-\frac{T}{t_2}} \Big). \quad (S3)$$

Here we have an analytical expression for the CW Mandel Q parameter for an ideal $(g^{(2)}(0) = 0)$ single photon emitter including bunching.

We can plot this function using values from CW $g^{(2)}(\tau)$ measurements under 250 µW excitation. In Fig. S8 the analytical $Q(T)$ expression is plotted for bunching parameter $a = 0.3$, antibunching and bunching times $t_1 = 2.7$ ns and $t_2 = 200$ ns, and single photon count rate $\langle I \rangle = 34$ kHz. Note that the count rate $\langle I \rangle$ already includes the total detection efficiency so accounts for losses in the optical path.

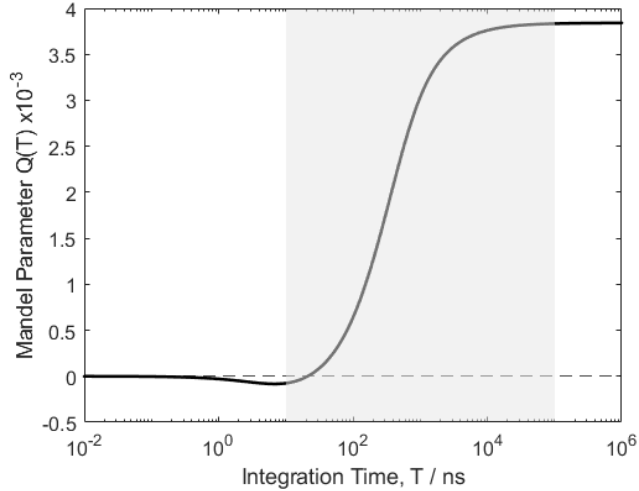

**Fig.S8.** Analytical solution for CW Mandel Q parameter. The function in Eq. (S3) was plotted with parameters $a = 0.3$, $t_1 = 2.7$ ns, $t_2 = 200$ ns, and single photon count rate $\langle I \rangle = 34$ kHz. Shaded region shows the range plotted for experimental and simulated data in Fig. 4 of the main text.

The shaded region indicates the range plotted for experimental and simulated $Q(T)$ in Fig. 4 of the main text. The limiting behaviour at high and low $T$ is the same as that in the experimental data; in particular $Q(T)$ tends to zero at low $T$. We also see that the crossover time from negative to positive $Q(T)$ occurs at around 21 ns, much lower than the 100 ns seen in the experimental data. The analytical model does not include all transition lifetimes of the three-level emitter $\tau_{ij}$, or the effect of detector deadtime.

It would also be possible to produce an analytical solution for the pulsed Mandel Q parameter. However, this would be more challenging because the pulsed $g^{(2)}(\tau)$ histograms are noisier due to the relatively low photon count rate under pulsed excitation. As such there are fewer constraints when choosing a function to use in Eq. (S1), with parameters which model pulsed $g^{(2)}(\tau)$ well.